# Designable ductility of a nano-network from few-layered graphene bonded with benzene: A molecular dynamics study

Jiao Shi [1,2*], Jia-Long Zhang [1], Xin Li [1], Bo Song [1*]

*1 College of Water Resources and Architectural Engineering, Northwest A&F University, Yangling 712100, China*

*2 State Key Laboratory of Structural Analysis for Industrial Equipment, Dalian University of Technology, Dalian 116024, China*

*Corresponding author's email address: shijiaoau@163.com (J. Shi); songbo943@163.com (B. Song)

**Abstract:** In nanoscale, motion operation of a nano-objective is usually realized by displacement load, which put forwards high requirement for ductility of material. Since pristine graphene has low ductility, once the stretching strain exceeds its critical value, it breaks in brittle style and loses ability to bear the external load quickly. Herein, to improve the ductility, a corrugated sandwich carbon nano-network model based on few-layered graphene is proposed, in which the two surface layers are bonded with several corrugated core layers via benzene molecules. Effects of factors such as the geometry of the carbon network, temperature, and strain rate, on the ductility are evaluated by molecular dynamics simulations. Conclusions are drawn for potential application of the new two-dimensional material with designable ductility.

## 1 Introduction

In evaluating the structural functions of a structure or device, we should first consider the material properties and material layout in the structure/device. With clear understanding of the material properties and appropriate layout, new materials with excellent properties can be developed. To form new nanomaterials with innovative physical properties, low-dimensional materials, such as carbon nanotubes (CNT), graphene, black phosphorene, WoS2, *h*-BN, etc, are candidate materials for design. Especially, benefiting from their excellent mechanical, electrical and thermal properties [1-5], two typical $sp^2$ carbon materials, i.e., graphene [6] and CNT [7], are widely applied in developing new carbon networks for multi-purposes. For example, Baughman and Galvão [8] proposed a new carbon network with auxetic property. Hall *et al.* [9] mixed single-walled CNTs and multi-walled CNTs to form buckypaper. Using graphene and CNTs, a three dimensional network has new mechanical and thermal properties was built by Xu *et al.* [10]. Cai *et al.* [11] evaluated the size and surface effects on the mechanical properties of a re-entrant two dimensional graphene-based kirigami [12]. Jiang *et al.* [13] introduced a new carbon allotrope called twin-graphene, which has semi-conductive with strain-dependent bandgap. In the reports of Liu and his colleagues [14, 15], irradiation methods are adopted to form carbon networks from multi-layered graphene. With shot peening method, Shi *et al.* [16] suggested to form a new carbon network from few-layered graphene, in which the bending stiffness was improved. By bonding two twisted layers of graphene, Chen *et al.* [17] formed a nanocomposite superstructure. Yang *et al.* [18, 19] proposed models of nanotextures from graphene ribbons, which can be applied in design of sensors. In the work published by Shi *et al.* [20], non-passivated diamondene [21, 22] was considered as a new carbon network, and its thermal and mechanical properties were evaluated. Meanwhile, Cai and his colleagues focused on developing nanotubes from diamondene [23-25].

At the nanoscale, motion and/or deformation of an object is commonly realized by a probe, which drives the object to move along a path via van der Waals force. For example, to estimate the frictional force between neighbor concentric CNTs, the inner tubes are generally pulled out of the outer tubes [26, 27] by probes. Stretching by probe tip, graphene kirigami can deform with large displacement [12]. In terms of mechanics, this kind of operation is called displacement load. Different to that under force load, the sample under displacement load deforms with the motion of the probe tip. In particular, the magnitude of stretching force varies during this period. Hence, when manipulating an object in nanoscale, displacement load is easier to implement than force load.

It is known that ductility is one of the most important mechanical properties of a material, which indeed illustrates its capacity of plastic deformation before rupture under tension. However, researches showed that the critical strains of pristine graphene ribbon along the armchair/zigzag directions are about 21%/14% at room temperature, respectively [16]. Once the stretching strain exceeds the critical value, it breaks in brittle style. In other words, the pristine graphene has low ductility, and thus hinders its application. As aforementioned, governing the layout of the materials is another powerful way to improve the properties of structure/device. This can be realized by by methods such as topology optimization approaches [28-33]. Typical lightweight composites like sandwich beams/plates/shells have already been widely used in aerospace engineering. In addition, these lightweight materials/structures can be fabricated by bonding surface layers and core layers together [34-36].

Here, to improve the ductility of pristine graphene, inspired by the idea of sandwich-like structures, we proposed a new sandwich-like graphene/benzene based carbon network. In the new carbon structure, the surface layers and core layers of the few-layered graphene ribbon are bonded with benzene molecules. More importantly, the mechanical behavior of the new carbon network is evaluated by molecular dynamics simulations.

## 2 Models and methodology

### 2.1 Models

The initial configurations of the two carbon networks are shown in Figure 1, in which the core has two or three layers respectively. Different from macro sandwich beam formed by adhesives [34], the surface layers and the core layers are bonded with benzene molecules in the present nano model.

In x-direction (stretching direction), the core has several unit cells, e.g., $N_p$=3. The initial length of surfaces equals $N_p \times L_p$. To evaluate the effect of $N_p$ on the strength of the network in x-direction, some other values of $N_p$ will be considered. More details on the involved models are listed in Table 1.

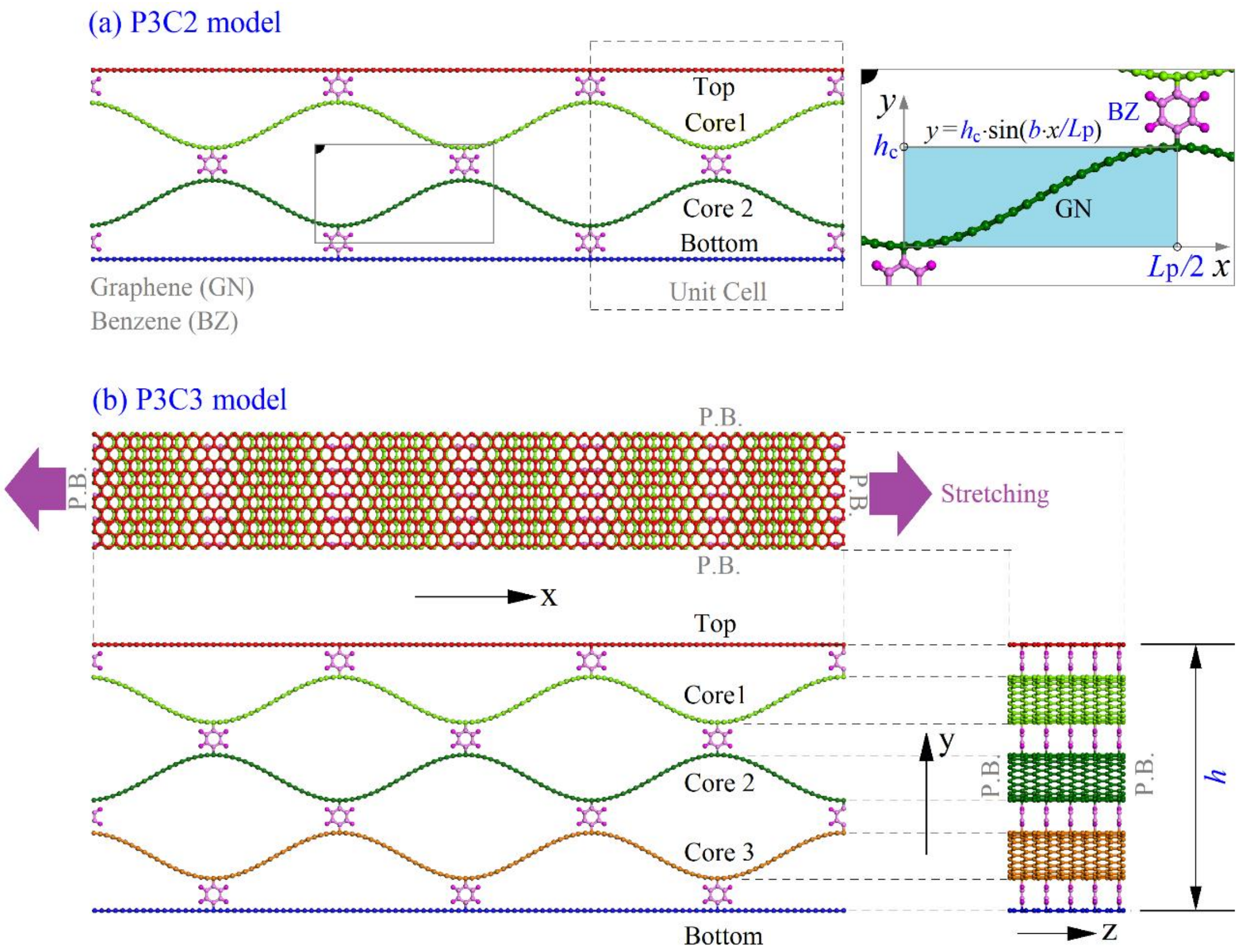

Figure 1 Initial geometries of sandwich nano-networks from graphene ribbons (GNs) bonded with benzene (BZ) molecules. The network contains two surface layers, i.e., Top and Bottom, and two or three core layers. The neighboring layers are connected with benzenes. The boundaries in x-/z-directions are set as periodic boundaries (P.B.). For simplicity, the core layers have the same shape of sine functions. (a) P3C2 model with two core layers ($N_c$=2) and three periodic cells ($N_p$=3), (b) Three orthographic views of P3C3 model with three core layers ($N_c$=3) and three periods ($N_p$=3) in stretching direction.

Table 1 Parameters of carbon networks. Note that "1080C" means the layer contains 1080 carbon atoms (excluding those in benzenes). "$R_{Llayer}$" is the ratio of the core layer length to the surface layer length.

| Model | $N_c$ | $N_p$ | $L_p$ (Å) | $h_c$ (Å) | $h$ (Å) | Surface layer | Core layer | $R_{Llayer}$ |
|---|---|---|---|---|---|---|---|---|
| P3C2 | 2 | 3 | 44.28 | 8.0 | 33.04 | 1080C | 1200C | 111.1% |
| P3C3 | 3 | 3 | 44.28 | 8.0 | 46.72 | 1080C | 1200C | 111.1% |
| P4C2 | 2 | 4 | 31.98 | 8.0 | 33.04 | 1040C | 1200C | 115.4% |
| P4C3 | 3 | 4 | 31.98 | 8.0 | 46.72 | 1040C | 1200C | 115.4% |
| P5C2 | 2 | 5 | 24.60 | 8.0 | 33.04 | 1000C | 1200C | 120.0% |
| P5C3 | 3 | 5 | 24.60 | 8.0 | 46.72 | 1000C | 1200C | 120.0% |
| P6C2 | 2 | 6 | 19.68 | 8.0 | 33.04 | 960C | 1200C | 125.0% |
| P6C3 | 3 | 6 | 19.68 | 8.0 | 46.72 | 960C | 1200C | 125.0% |

## 2.2 Methodology

### 2.2.1 Flowchart of molecular dynamics simulation

To reveal the tensile properties of the carbon networks from graphene bonded with benzenes, classical molecular dynamics simulation approach is used in this study. The simulations are conducted on the large-scale atomic/molecular massively parallel simulator (LAMMPS) [37]. In a simulation, the interactions among atoms are illustrated by the adaptive intermolecular reactive empirical bond order (AIREBO) potential [38], which is popular in describing the bonding and non-bonding interactions between atoms in a hydrocarbon system.

$$\begin{cases} P = P_{\mathrm{REBO}} + P_{\mathrm{Torsion}} + P_{\mathrm{L\text{-}J}} \\ P_{\mathrm{REBO}} = \sum_{i} \sum_{j(j>i)} \left[ V_{ij}^{R}(r_{ij}) - b_{ij} V_{ij}^{A}(r_{ij}) \right] \\ P_{\mathrm{Torsion}} = \frac{1}{2} \sum_{i} \sum_{j(j\neq i)} \sum_{k(k\neq i,j)} \sum_{l(l\neq i,j,k)} w(r_{ij}) \cdot w(r_{ij}) \cdot w(r_{ij}) \cdot V_{\mathrm{Torsion}}(\omega_{ijkl}) \\ P_{\mathrm{L\text{-}J}} = \sum_{i} \sum_{j(j>i)} 4\varepsilon \left[ \left( \frac{\sigma}{r_{ij}} \right)^{12} - \left( \frac{\sigma}{r_{ij}} \right)^{6} \right] \end{cases}, \tag{1}$$

where $P_{\mathrm{REBO}}$ represents the short-range REBO potential [39]. $V_{ij}^{R}$ and $V_{ij}^{A}$ are repulsive and attractive pairwise potentials that determined by the atom types of atoms $i$ and $j$. $r_{ij}$ is the bonded atom distance with $b_{ij}$ as the many-body term. The cutoff of bond is set as 0.2 nm [40]. $P_{\mathrm{Torsion}}$ depends on the dihedral angle $\omega$ of the atoms $i$, $j$, $k$, and $l$, with bond weight $w_{ij}$ is in [0, 1]. $V_{\mathrm{Torsion}}$ is the dihedral-angle potential. $P_{\mathrm{L\text{-}J}}$ describes the non-bonding interactions (Lennard-Jones interaction [41]) with $\sigma_{\mathrm{C\text{-}C}}$=0.34 nm, $\sigma_{\mathrm{H\text{-}H}}$ =0.265 nm, $\sigma_{\mathrm{C\text{-}H}}$=0.3025 nm, $\varepsilon_{\mathrm{C\text{-}C}}$=2.84 meV, $\varepsilon_{\mathrm{H\text{-}H}}$=1.5 meV, $\varepsilon_{\mathrm{C\text{-}H}}$=2.065 meV, and the cutoff is 1.02 nm.

For showing the deformation of a sample accurately, the time step for integration of atoms' motion equations is set as 0.001 picosecond (ps). In stretching along the x-direction, the sample is in a simulation box at an NPT (constant Number of atoms, constant Pressure and Temperature of system) ensemble with no pressure on the other four sides, and temperature is controlled by Nose-Hoover thermostat [42, 43]. Before stretching, the sample is relaxed for 10 ps. After that, deform the simulation box with the specified strain rate, and then remap the coordinates of the atoms every picosecond.

### 2.2.2 Features of the new carbon network

For showing the tension state, variation of potential energy (VPE) per atom will be recorded, which is the difference between the "current" and the "initial" potential per atom of the system in stretching, i.e.,

$$\mathrm{VPE} = [P(t) - P(t_0)]/N, \tag{2}$$

Where $P(t)$ and $P(t_0)$ are the potential energy of the system at moments $t$ and $t_0$, respectively. They can be obtained by substituting the positions of atoms in Eq.(1). "$N$" is the number of carbon atoms in the system.

The stretching strength of a sample will be expressed by the critical values of x-strain, and the effects of the major factors, such as temperature, strain rate/loading speed, $N_{\mathrm{p}}$ and $N_{\mathrm{c}}$, on the strength will be demonstrated in numerical tests. Accordingly, the stress along the stretching direction, e.g., x-stress, will be written down at each x-strain (the engineering strain rate times steps in x direction). Actually, the x-stress is the virial stress [44] with considering 3.4 Å as the thickness of each surface layer. The normal stress in x-direction in the virial stress tensor is defined for

$$\text{x-Stress} = \frac{1}{V} \sum_{k\in\Omega} \left( -m^{(k)} \left( u_{\mathrm{x}}^{(k)} - \bar{u}_{\mathrm{x}} \right)^{2} + \frac{1}{2} \sum_{l\in\Omega} \left( u_{\mathrm{x}}^{(l)} - u_{\mathrm{x}}^{(k)} \right) f_{\mathrm{x}}^{(kl)} \right), \tag{3}$$

where $k$ and $l$ are the $k$-th and $l$-th atoms in the domain Ω with volume of $V$. $m(k)$ is the mass of atom $k$. with $u^{(k)}_{\mathrm{x}}$ as the x-component of its velocity vector. $\bar{u}_x$ is the speed of the mass center of the system in x-direction. $f_x^{(kl)}$ is the x-component of the force on atom $k$ by atom $l$.

When stretching under given engineering strain rate of $\dot{\varepsilon}$, the length of sample can be calculated by

$$L(t) = L(t_0) \times \left[1 + \dot{\varepsilon}(t - t_0)\right], \tag{4}$$

where, $L(t_0)$ is the initial length at moment $t_0$, $\dot{\varepsilon}$ $(t - t_0)$ is the engineering strain.

## 3. Results and discussion

3.1 Responses of the carbon networks with $N_p$=3 under uni-axial tension

*(a) P3C2 model*

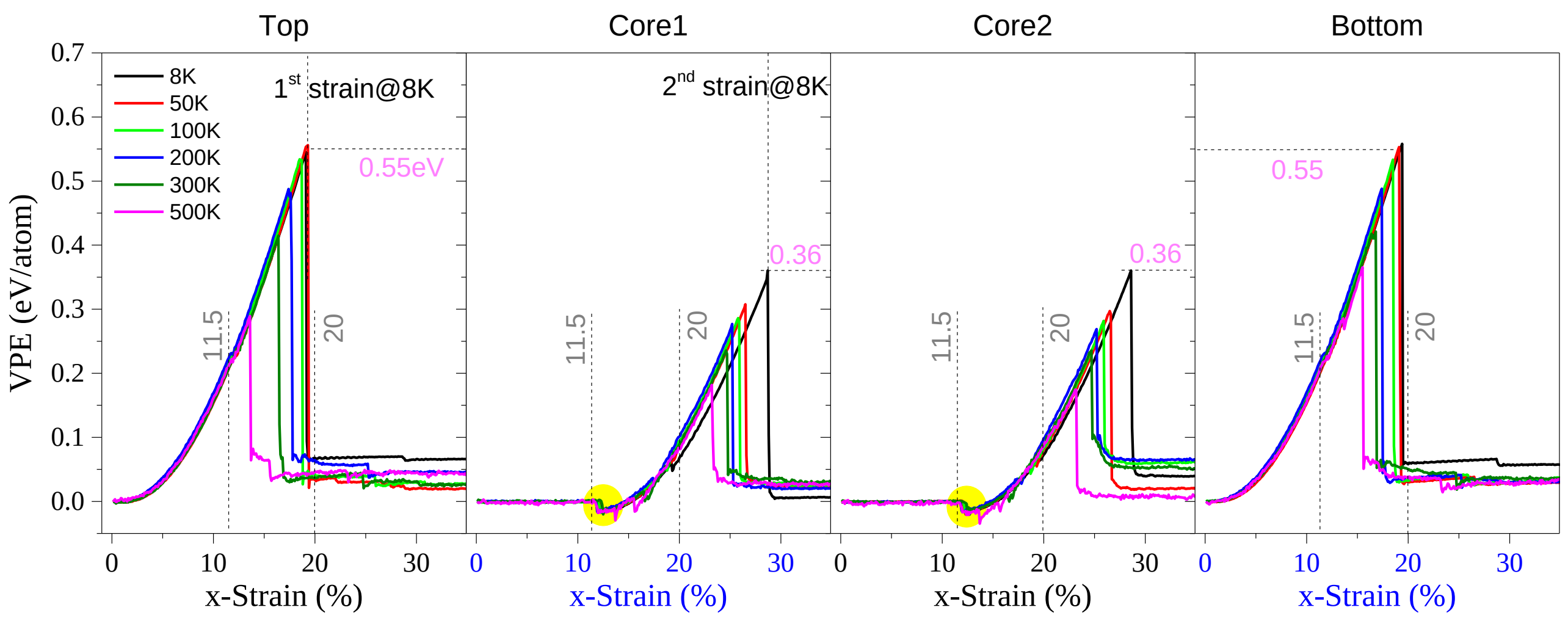

Figure 2 Histories of VPE per atom of the P3C2 sample under the strain rate of 0.1%/ps at different temperatures. Note that 1st strain=1st $\varepsilon_c$, and 2nd strain=2nd $\varepsilon_c$.

In Figure 2, the VPE of the four layers (i.e., top, core1, core2 and bottom) of the P3C2 sample at temperature between 8 K and 500 K are illustrated. The results indicate several characteristics of the sample under uni-axial tension.

First, the two surface layers have approximately the same critical strain, i.e., **1st critical strain** labeled as 1st $\varepsilon_c$. Similarly, the two core layers have the same **2nd critical strain** (and labeled as 2nd $\varepsilon_c$). The two critical values demonstrate the stretching strength of the sample. Details on ductility will be given below.

Second, temperature influences the tensile strength significantly. For example, at temperature below 100 K, the two surface layers have the critical strain of ~19%. However, the value of 1st $\varepsilon_c$ reduces to 16.4% at 300 K or 13.6% at 500 K, which is lower than 21% of the pristine graphene at 300 K [16]. This reduction of strength is due to the existence of $sp^3$ bonds that connect surface layers and core layers via benzene molecules.

Third, the VPE curve of a surface layer contains two stages, i.e., continuous increasing and sharp drop after the strain exceeds the 1st $\varepsilon_c$. However, the VPE curve of core layer has one more stage, i.e., VPE keeps unchanged near zero with small x-strain, which demonstrates that the core layer has no elastic deformation in this stage.

Fourth, for the sample with x-strain between 11.5% and 20%, VPE of the core layers decreases first and then increases till approaching 1st $\varepsilon_c$. The decrease of VPE is due to two facts. **One** is the core layers being attracted by and attaching to the surface layer (Figure 3a), and **the other** is the core layers attach together during deformation (12.3%-snapshot in Figure 3a). In this period, the core buckles [23, 45-47] and the carbon network shrinks along y-direction according to the snapshot in Figure 3a. After that, the core layers work together with the two surface layers against stretching load, which means that the system is in co-working state. For example, at 8 K, the carbon network keeps undamaged before x-strain approaching 19%. However, at 19.1% (1st $\varepsilon_c$), the surface layer breaks (Figure 3b). The core layers break till x-strain reaches 28.6% (2nd $\varepsilon_c$) (Figure 3c and Table 2), which is greater than that of the pristine graphene at the same temperature. It means that the ductility of the carbon network is enhanced when considering the bond breakage in surface layers as "plastic deformation".

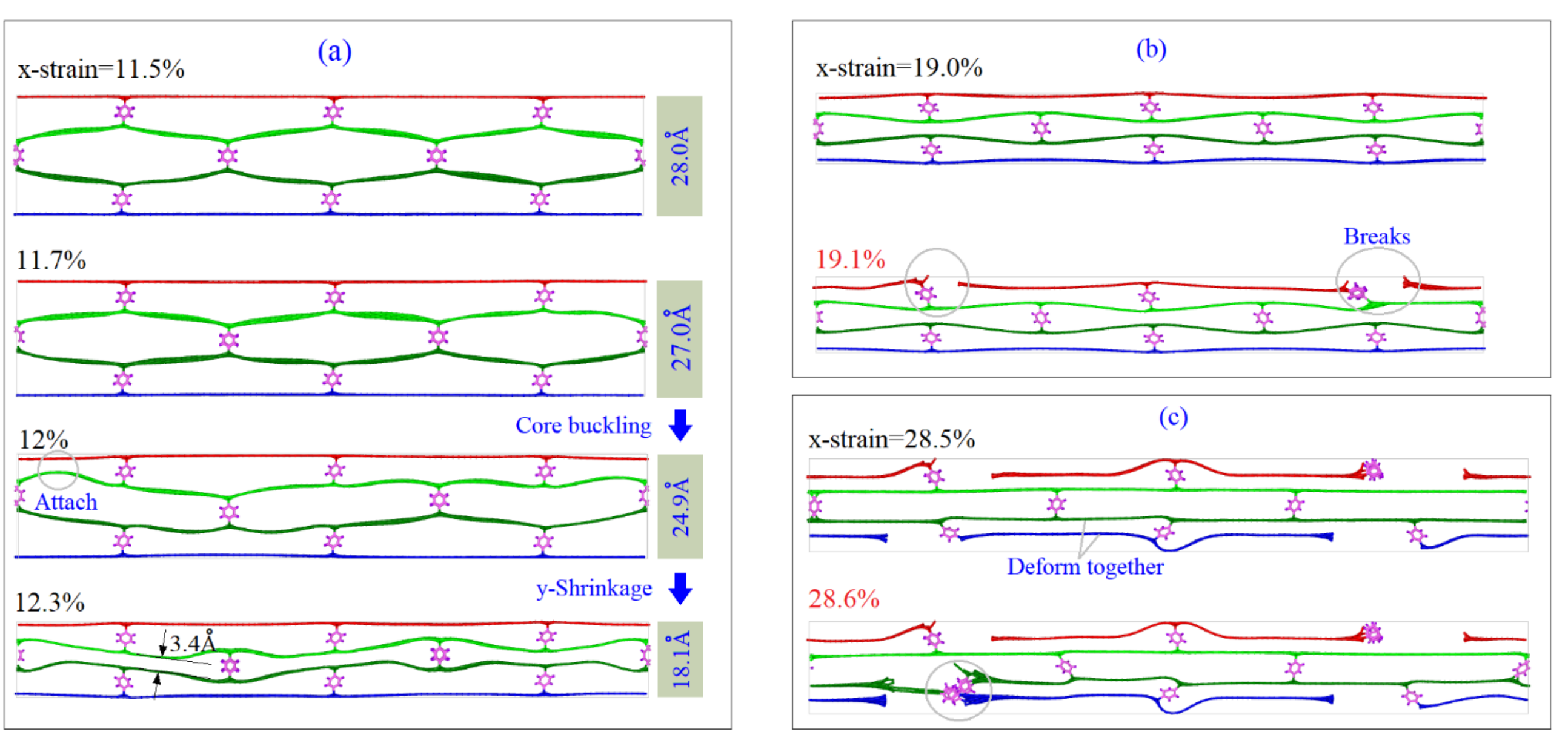

Figure 3 Typical snapshots of the P3C2 sample under the strain rate of 0.1%/ps at 8 K. (a) the sample buckles with the height decreasing sharply when the x-strain is between 11.7% and 12.3%, (b) top layer breaks at x-strain=19.1%, and (c) Core2 breaks when x-strain=28.6%.

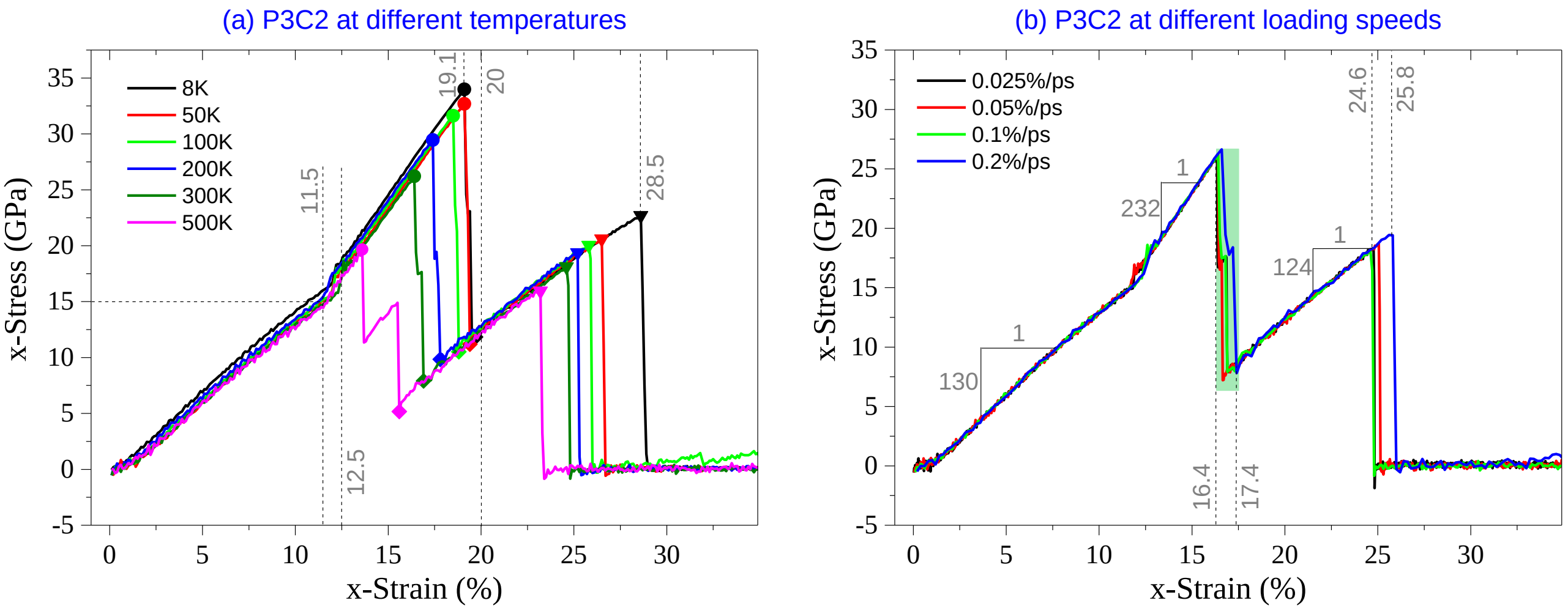

Figure 4 Stress-strain curves of the P3C2 sample under tension. (a) at different temperatures with strain rate of 0.001%/ps, (b) at different loading speeds with temperature of 300 K.

In Figure 4, we show the stress-strain curves of the P3C2 sample that under different stretching conditions. In Figure 4a, the curves differ slightly before x-strain approaching 11.5%, and the **(first) slope/modulus** is ~130 GPa (Figure 4b). When x-strain is between 12.5% and 19.1%, the increment of stress is ~15.3 GPa, which means the **(second) modulus** is ~232 GPa. The second modulus of sample is much higher than the first modulus due to two facts, i.e., the four layers work together in the co-working period, and the sample shrinks in y-direction leading to lower volume of sample (see Eq. (3)). For the sample with x-strain between 20% and 28.5%, the stress increases 10.5 GPa, which indicates that the **(third) modulus** of the sample is ~124 GPa. It is lower than the first modulus only because the volume of simulation box becomes higher when the surface layers break (Figure 3c).

From Eq. (3), we know that the magnitude of stress is inversely proportional to the volume of simulation box. To reflect the stretching strength of the carbon network, we measure of strength with two critical strains. As shown in Table 2, the 1st $\varepsilon_c$ decreases with increasing of temperature, and the 2nd $\varepsilon_c$ changes from 28.6% at 8 K to 23.2% at 500K. However, after breaking of surface layers, the increment of strain is not sensitive to temperature.

In Figure 4b, the stress-strain curves with respect to different loading speeds are shown. The curves are the same when x-strain is between 0 and 11.5% or between 13.2% and 14.6%, or between 18% and 24.6%. It says that the three moduli are the same even if the stretching speed changes from 0.025%/ps to 0.2%/ps. Obvious differences exist among the curves when x-strain is between 11.5% and 13.2% or between 16.4% and 17.4%. When x-strain is in the interval [11.5%, 13.2%), the core layers has different buckling process at different loading speeds. When in the interval [16.4%, 17.4%), the surface layers have different breakage process.

Table 2 Typical values of stress in stretching direction of the P3C2 sample with strain rate of 0.1%/ps at different temperatures. "$\sigma$" is x-stress, and "ε" is x-strain. Incremental strain $\Delta\varepsilon$=2nd $\varepsilon_c$ - $\varepsilon_{Tr}$.

| T | ●Peak 1 | | | ♦Trough | | ▾Peak 2 | | |
|---|---|---|---|---|---|---|---|---|
| | 1st $\sigma$/GPa | 1st $\varepsilon_c$ | ↓$\Delta\sigma$/GPa | $\sigma$/GPa | $\varepsilon_{Tr}$ | 2nd $\sigma$/GPa | $\Delta\varepsilon$ | 2nd $\varepsilon_c$ |
| 8 K | 33.98 | 19.1% | 22.79 | 11.19 | 20.4% | 22.70 | 8.2% | 28.6% |
| 50 K | 32.69 | 19.1% | 20.85 | 11.84 | 18.7% | 20.61 | 7.8% | 26.5% |
| 100 K | 31.62 | 18.5% | 21.10 | 10.52 | 18.8% | 20.03 | 7.0% | 25.8% |
| 200 K | 29.46 | 17.4% | 19.62 | 9.84 | 17.8% | 19.38 | 7.4% | 25.2% |
| 300 K | 26.22 | 16.4% | 18.31 | 7.91 | 16.9% | 18.12 | 7.7% | 24.6% |
| 500 K | 19.66 | 13.6% | 14.49 | 5.17 | 15.6% | 15.95 | 7.6% | 23.2% |

*(b) P3C3 model*

In Figure 5a, the VPE curves of the five layers in P3C3 model under the strain rate of 0.1%/ps at 300 K are given. The two surface layers have the same values of VPE before breakage happens. The top layer breaks at x-strain=16.7% , while the bottom layer breaks when x-strain exceeds 17.4%. The three core layers have approximately the same VPE in tension before breakage happening at x-strain=25.8% (Figure 5b). The co-working region of x-strain is between 12.3% and 16.6%.

The same phenomenon, i.e., VPE of the core layers decreases soon after x-strain > 11.9%, can also be observed in Figure 5a. According to the snapshots in Figure 5b, the sample shrinks sharply between x-strain=11.9% and 12.3%. The core structure buckles [23, 45-47] and the three core layers attach together. Between 16.6% and 17.4% of x-strain, VPE of the core layers has fluctuation, which is caused by the co-working of the surface layers and the core1 and core3 layers in deformation (similar to the 25.7%-snapshot in Figure 5b).

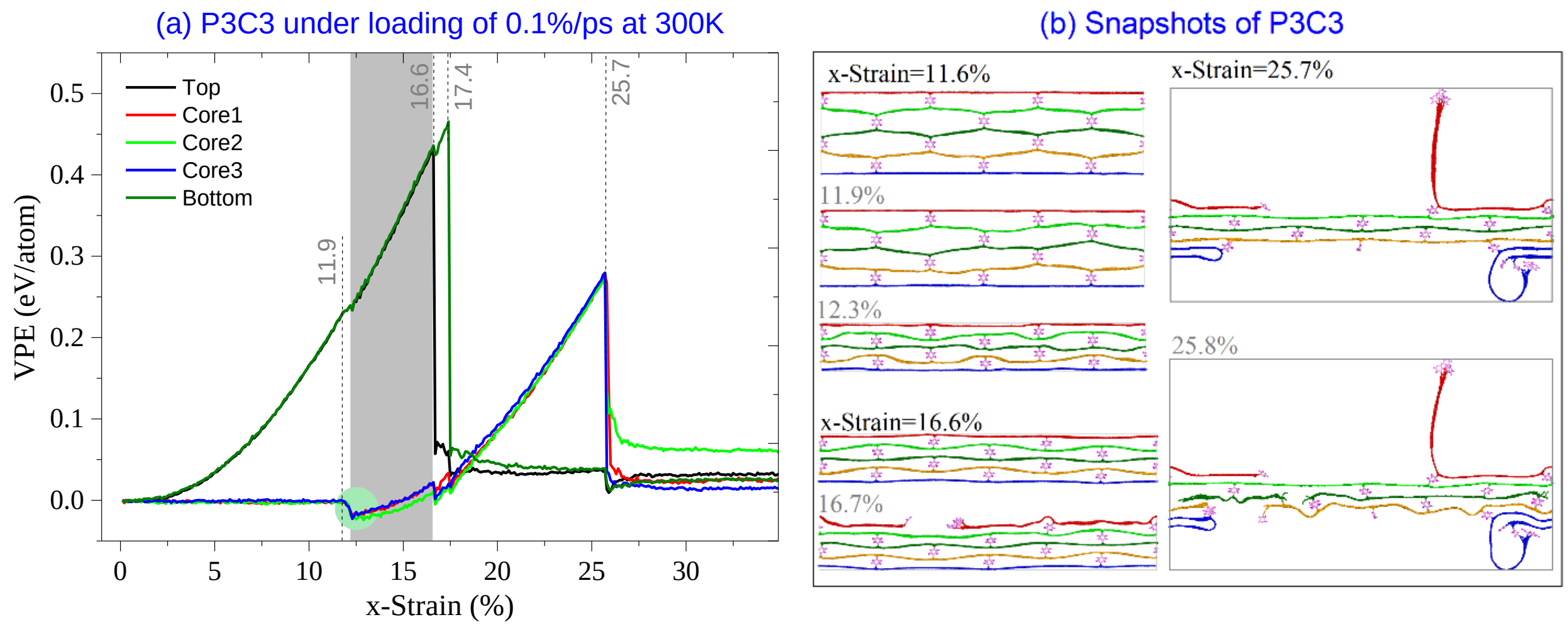

Figure 5 VPE per atom of the P3C3 sample under the strain rate of 0.1%/ps at 300 K. (a) VPE curves, (b) snapshots.

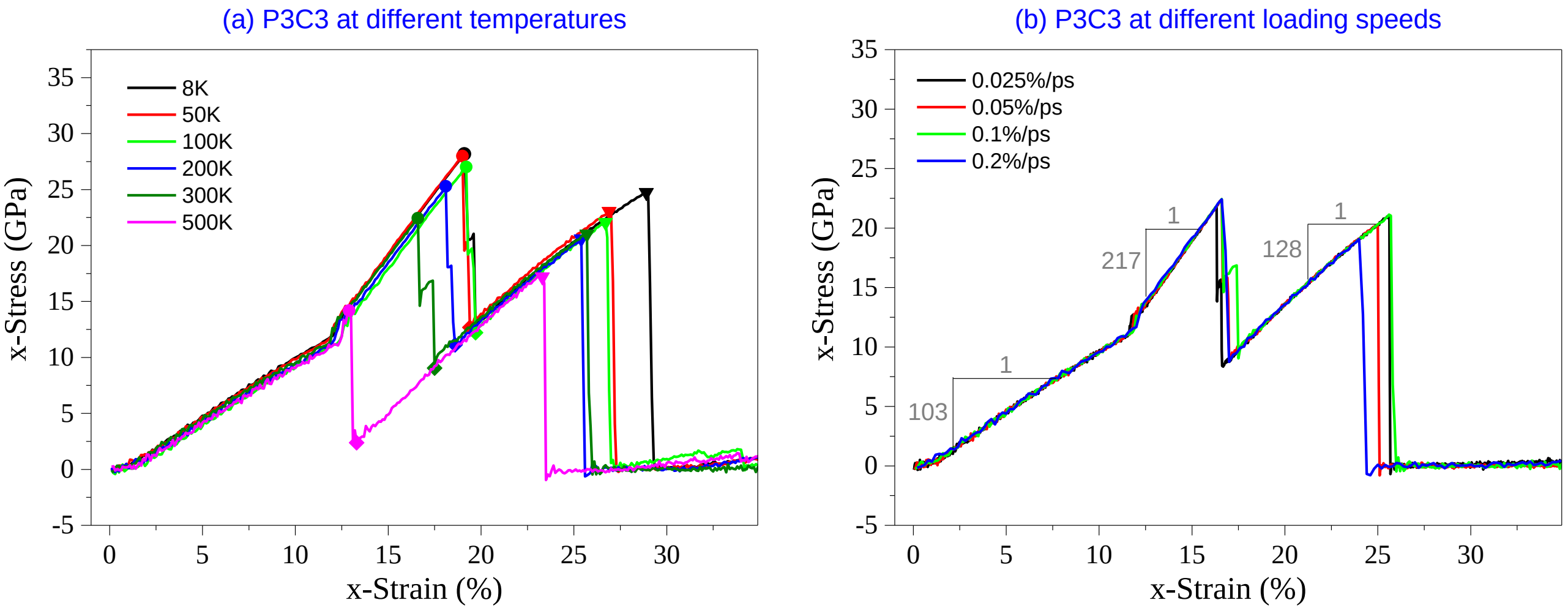

Figure 6 Stress-strain curves of the P3C3 sample under tension. (a) at different temperatures with strain rate of 0.1%/ps, (b) at different loading speeds with temperature of 300 K.

Besides, the stress-strain curves of the P3C3 sample with considering the effects of temperature and loading rate are given in Figure 6. Obviously, the critical strains are sensitive to temperature, but not sensitive to the loading rate. According to the results listed in Table 3, the critical strains are slightly different to those of the P3C2 model at the same temperature (except at 500 K). It means that the critical strain is independent of the number of core layers.

However, as comparing to P3C2 at the same temperature, P3C3 has lower stress level due to larger volume with the same critical strain according to Eq.(3). Comparing the values of sharp jump of stress ($\downarrow\sigma$) at the first critical strains in Table 2 and Table 3, P3C3 has lower stress decline than P3C2, i.e., the stress fluctuation of P3C3 is weaker. It implies that more core layers will reduce fluctuation of stress, which is significant for application of the carbon network in bearing displacement load.

We also calculate the three moduli of P3C3 according to its stress-strain curves. It shows that the **first modulus** is only 103 GPa (Figure 6b), which is less than 130 GPa of P3C2. In this stage, only surface layers bear the external load. The two samples have the same surface layers, but P3C3 has larger volume that leads to lower stress level. Hence, P3C3 has lower 1$^{st}$ modulus. **The second and the third moduli** are 217 GPa and 128 GPa, respectively, which are slightly different from those of P3C2. In this stage, only the core layers are under stretching, and the sample's volume is approximately proportional to the number of core layers. Therefore, the 2nd or the 3rd moduli of the two samples differ slightly. **It is also the major reason for not providing the results of the samples with more core layers in this study**.

Table 3 Typical values of stress in stretching direction of the P3C3 sample with strain rate of 0.1%/ps at different temperatures. "$\sigma$" is x-stress, and "ε" is x-strain. Incremental strain $\Delta\varepsilon = 2^{nd}\ \varepsilon_c - \varepsilon_{Tr}$.

| T | ●Peak 1 | | | ♦Trough | | ▼Peak 2 | | |
|---|---|---|---|---|---|---|---|---|
| | $1^{st}\ \sigma$/GPa | $1^{st}\ \varepsilon_c$ | ↓$\Delta\sigma$/GPa | $\sigma$/GPa | $\varepsilon_{Tr}$ | $2^{nd}\ \sigma$/GPa | $\Delta\varepsilon$ | $2^{nd}\ \varepsilon_c$ |
| 8 K | 28.2 | 19.1% | 15.5 | 12.6 | 19.7% | 24.7 | 9.2% | 28.9% |
| 50 K | 28.0 | 19.0% | 15.3 | 12.7 | 19.4% | 23.1 | 7.8% | 26.9% |
| 100 K | 27.0 | 19.2% | 14.7 | 12.2 | 19.7% | 22.1 | 7.0% | 26.7% |
| 200 K | 25.3 | 18.1% | 14.5 | 11.1 | 18.6% | 20.6 | 7.4% | 25.4% |
| 300 K | 22.4 | 16.6% | 13.5 | 9.0 | 17.5% | 21.0 | 7.7% | 25.7% |
| 500 K | 14.2 | 12.9% | 11.8 | 2.4 | 13.3% | 17.2 | 7.6% | 23.3% |

3.2 Responses of carbon networks with more periodic cells in stretching direction

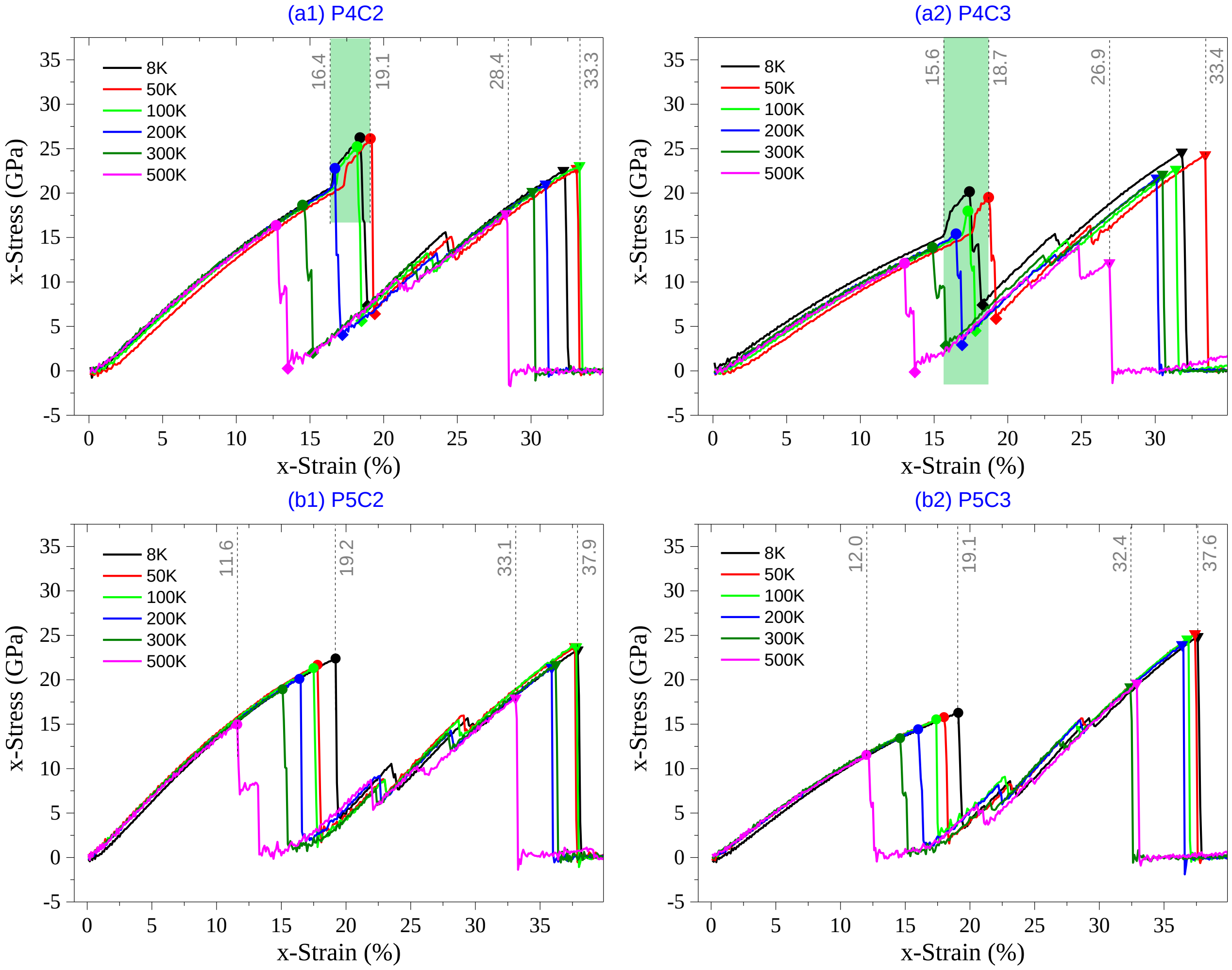

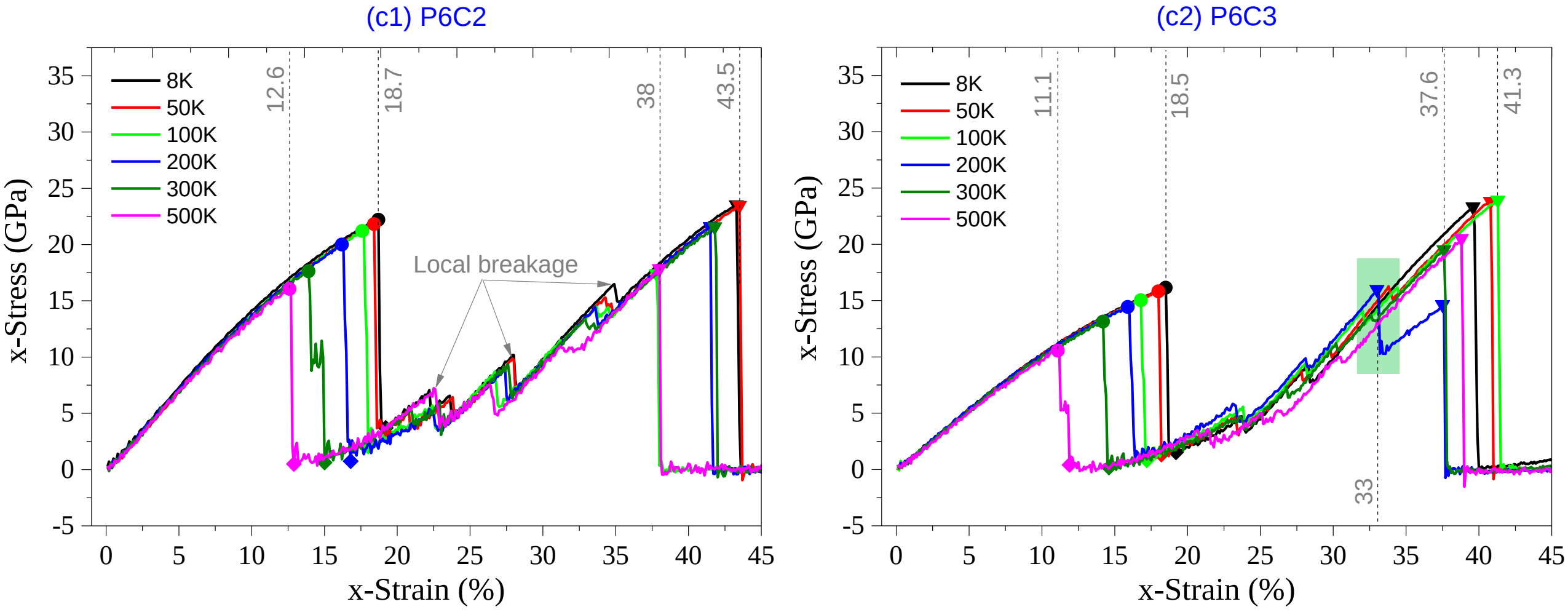

Figure 7 Stress-strain curves of the samples under loading speed of 0.1%/ps at different temperatures. (a) P4C2 & P4C3, (b) P5C2 & P5C3, and (c) P6C2 & P6C3.

According to the discussion in above section, the stress strain curve can well describe the mechanical behavior of the sample under uni-axial tension. In Figure 7, the stress-strain curves of the samples with more periodic cells are given with considering the temperature effect. One can find that the two samples with $N_p$=4, have a narrow co-working region at low temperature (in the light green rectangular in Figure 7a). However, when $N_p$>4, there is no co-working region of all the layers in a sample. The reason is that the length difference between the core and the surface layers, i.e., $N_{Llayer}$, becomes higher when $N_p$ is greater.

Another interesting phenomenon in Figure 7 is that more sudden jumps in the stress-strain curves happen after the surface layers break (i.e., x-strain>1$^{st}$ $\varepsilon_c$) when the sample has more periodic cells. The reason is that the surface breakage first happens at one benzene-bonded line in the surface layer (either top or bottom), which means that the broken surface layers and the core layers are still connected via the rest lines of benzene molecules (Figure 8b and Figure 9b). It also indicates **the ductility of the carbon network is improved when the local breakages are considered as plastic deformation of the sample**. During further stretching, the co-working segments in the surface layers break one by one (Figure 8b and Figure 9b), and each breakage leads to a sudden drop of stress. In further relaxation of system, VPE (Figure 8a and Figure 9a) decreases due to self-attraction between core layers.

When comparing the two samples with the same $N_p$ but different $N_c$, we find that the difference between critical strains is negligible at the same temperature. It further verifies that the number of core layers does not influence the critical strains.

When observing the sudden change of the configurations in Figure 8b, it can be found that the sample has no buckling of core layers before the surface layer breaks (comparing snapshots with respect to x-strain=14.9% and 15.8%). The reason is that the core layers are obviously longer than the surface layers. The core layer is still not bearing the external deformation when the surface layers break completely. Without heavy elongation, the core layers cannot get closer to each other, they also have no chance to attach together. After breakage of surface layers, the core layer shrinks quickly in transversal direction so as to bear the stretching deformation.

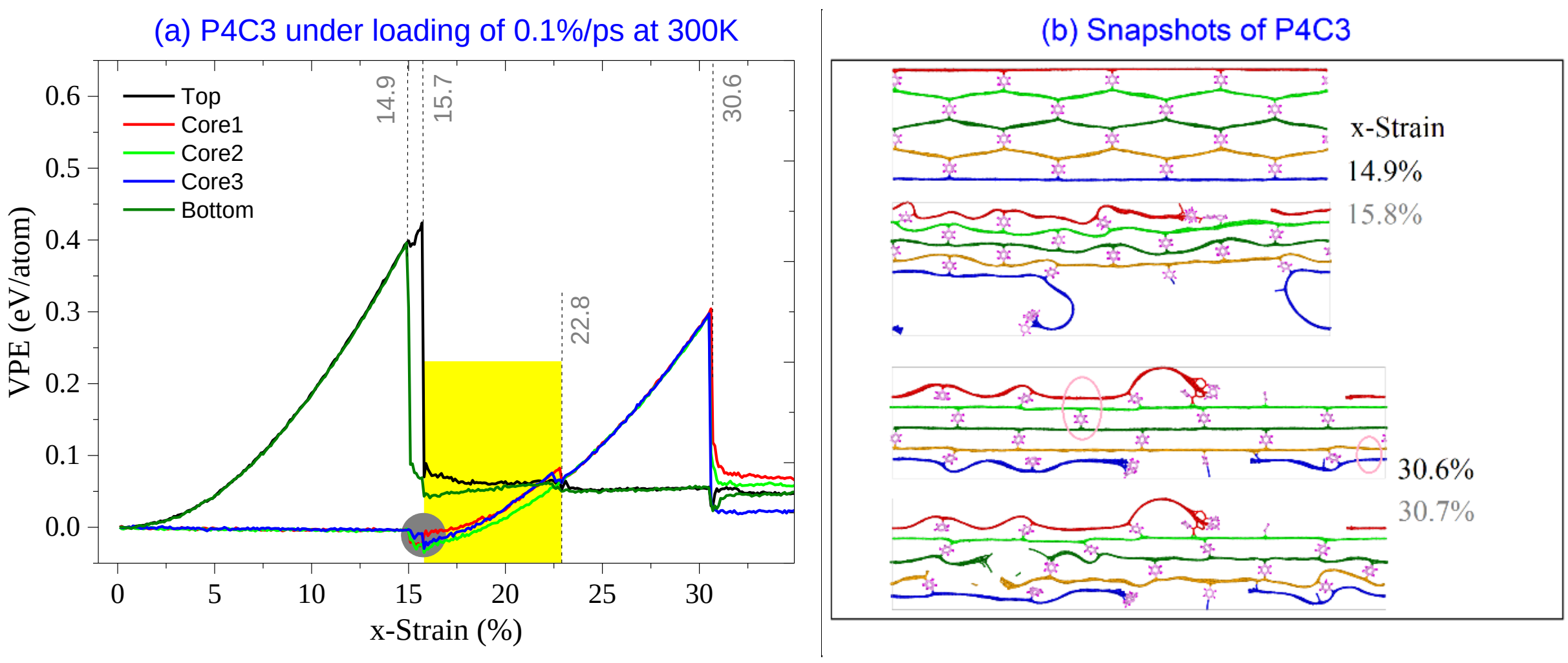

Figure 8 VPE curves of P4C3 under the loading speed of 0.1%/ps at 300 K. (a) VPE, (b) snapshots.

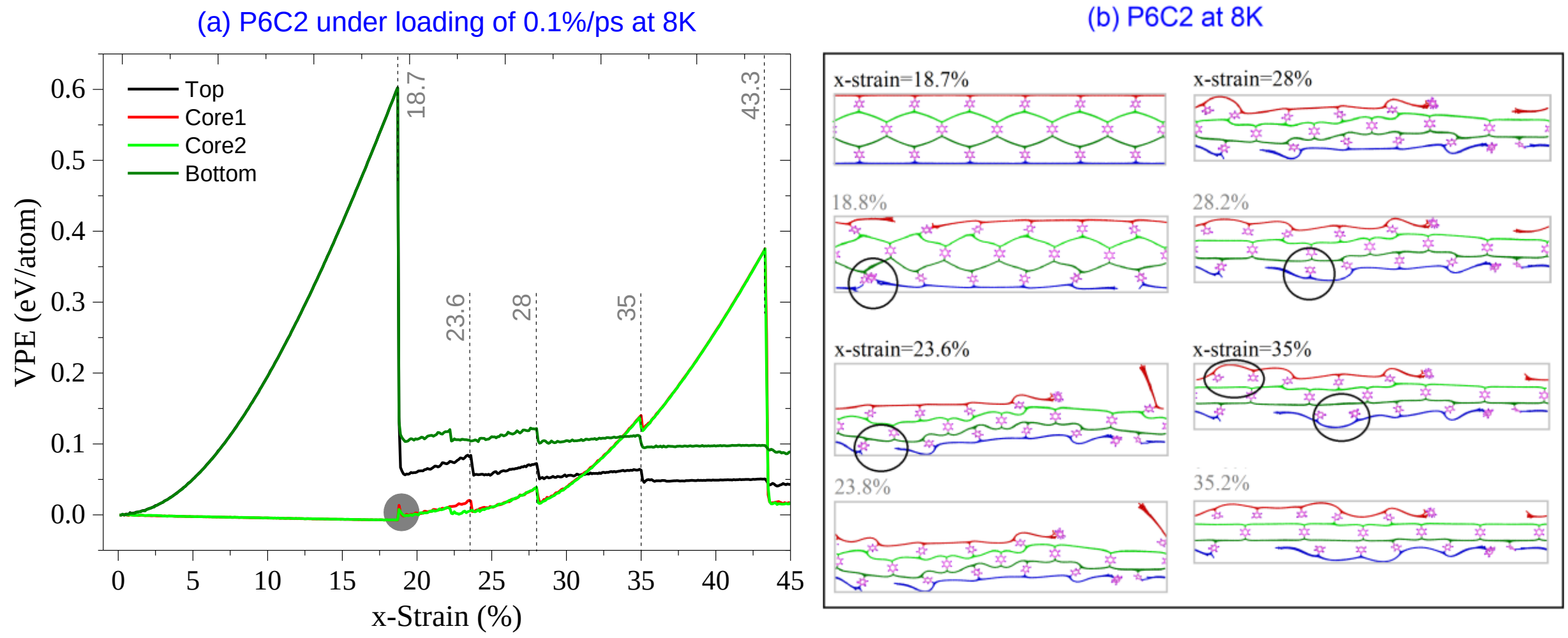

Figure 9 VPE curves of P6C2 under the loading speed of 0.1%/ps at 8 K. (a) VPE, (b) snapshots.

3.3 Analysis of critical strains with respect to $N_p$

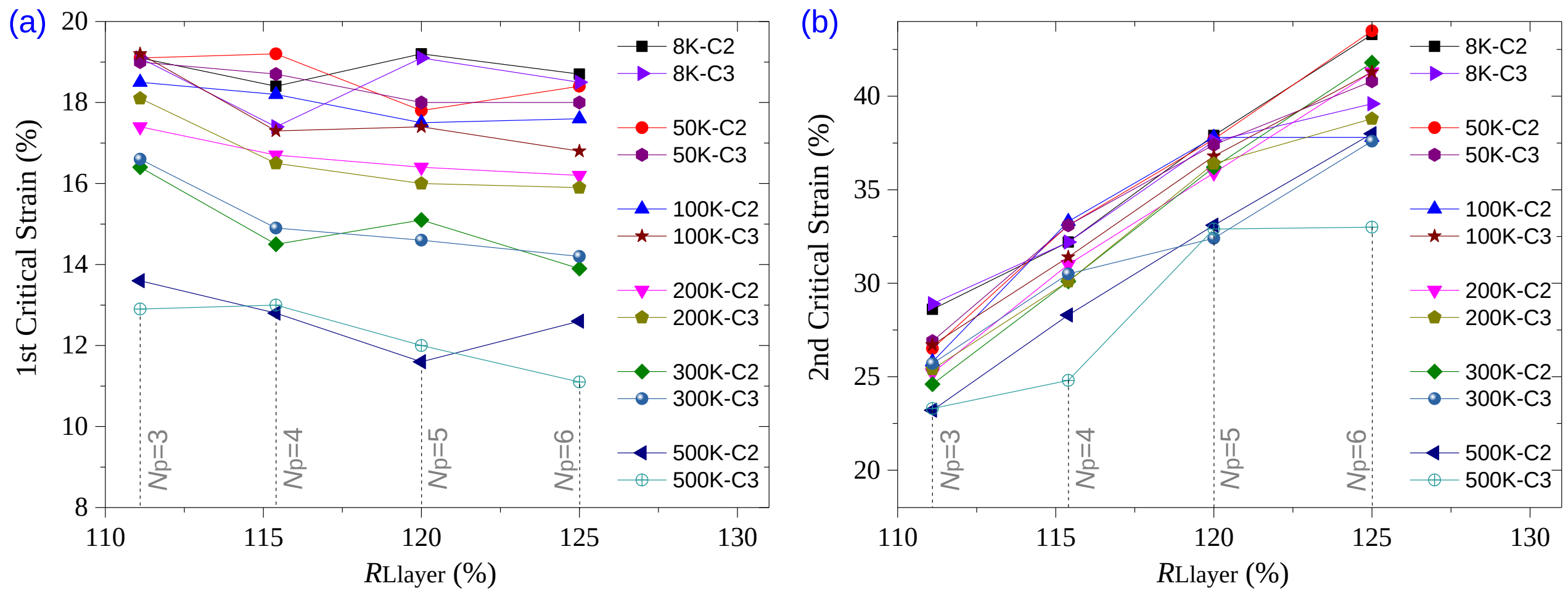

Figure 10 Critical strains of carbon networks under uniaxial load with strain rate of 0.1%/ps. (a) 1st critical strain *vs* $R_{\text{Llayer}}$, (b) 2nd critical strain *vs* $R_{\text{Llayer}}$.

In Figure 10, the critical strains of the carbon networks are illustrated with following characteristics, i.e., (1) the 1st critical strain decreases with increasing $N_p$. (2) The value of the 1st critical strain is lower at higher temperature. At the same temperature, the 1st critical strain is less than that of the pristine graphene. Whereas, (3) the 2nd critical strain is proportional to $N_p$. (4) the 2nd critical strain decreases with increasing temperature. However, at a specific temperature, it is greater than that of the pristine graphene. (5) The number of core layers has slight influence on the critical strain.

Since the 2nd critical strain increases linearly with $N_p$, a new carbon network with higher ductility can be designed with specific $N_p$, and this will benefit its application in bearing large deformation.

## 4 Conclusions

Pristine graphene breaks in a brittle manner when tensile strain reaches about 20% at room temperature. In this work, we design a carbon network with larger ductility by bonding graphene ribbons with benzene molecules. Using molecular dynamics simulation approach, the strength and ductility of the new carbon network is evaluated with consideration of the effects of cell geometries, temperature and loading speed. According to the results, the following conclusions are drawn.

First, due to the length difference between the surface and core layers, the carbon network under uniaxial load contains two critical strains. The surface layers break at the 1st critical strain, and the core layers break at the 2nd critical strain. At the same temperature, the 2nd critical strain is greater than that of the pristine graphene.

Second, when the length difference between surface and core layers is less than 20%, the stress-strain curve of the carbon network has three different moduli, i.e., the 1st modulus since only the surface layers bearing load, the 2nd modulus is the highest among the three moduli due to co-working of all the layers, and the 3rd modulus with only core layers bearing load.

Third, the critical strains of a carbon network decrease with temperature, but slightly depend on either loading speed or the number of core layers.

Final, the maximum critical strain with respect to breakage of the core layers depends on the length difference of the layers. Hence, the ductility of the carbon network is designable, which is significant for a material under large displacement load.

**Compliance with Ethical Standards:** The authors comply with the ethical rules for this journal.

**Credit authorship contribution statement: Jiao Shi:** Conceptualization, Formal analysis, Funding acquisition, Investigation, Methodology, Project administration, Supervision, Writing-original draft. **Jialong Zhang:** Data curation, Formal analysis, Investigation. **Xin Li:** Data curation, Formal analysis, Investigation. **Bo Song:** Formal analysis, Investigation, Methodology, Writing-review & editing, Visualization.

**Acknowledgement:** Financial supports from National Key Research and Development Plan, China (Grant No.: 2017YFC0405102) and State Key Laboratory of Structural Analysis for Industrial Equipment, Dalian University of Technology, Dalian 116024, China (Grant No.: GZ18111), are acknowledged.

**Conflict of Interests:** The authors declare that they have no conflict of interests.

## Appendix

**Table A1** Critical strains and stresses in stretching direction of the samples under the strain rate of 0.1%/ps at different temperatures. "$1^{st}$ $\sigma$" is first peak value of x-stress, and "$1^{st}$ $\varepsilon$" is the first critical value of x-strain with respect to breakage of surface layers. $2^{nd}$ $\varepsilon$ is the second critical value with respect to thorough breakage of all the layers.

| | P3C2 model | | | | P3C3 model | | | |
|---|---|---|---|---|---|---|---|---|
| | $1^{st}$ $\sigma$/GPa | $1^{st}$ $\varepsilon_c$ | $2^{nd}$ $\sigma$/GPa | $2^{nd}$ $\varepsilon_c$ | $1^{st}$ $\sigma$/GPa | $1^{st}$ $\varepsilon_c$ | $2^{nd}$ $\sigma$/GPa | $2^{nd}$ $\varepsilon_c$ |
| 8K | 33.98 | 19.1% | 22.70 | 28.6% | 28.2 | 19.1% | 24.7 | 28.9% |
| 50K | 32.69 | 19.1% | 20.61 | 26.5% | 28.0 | 19.0% | 23.1 | 26.9% |
| 100K | 31.62 | 18.5% | 20.03 | 25.8% | 27.0 | 19.2% | 22.1 | 26.7% |
| 200K | 29.46 | 17.4% | 19.38 | 25.2% | 25.3 | 18.1% | 20.6 | 25.4% |
| 300K | 26.22 | 16.4% | 18.12 | 24.6% | 22.4 | 16.6% | 21.0 | 25.7% |
| 500K | 19.66 | 13.6% | 15.95 | 23.2% | 14.2 | 12.9% | 17.2 | 23.3% |

| | P4C2 model | | | | P4C3 model | | | |
|---|---|---|---|---|---|---|---|---|
| | $1^{st}$ $\sigma$/GPa | $1^{st}$ $\varepsilon_c$ | $2^{nd}$ $\sigma$/GPa | $2^{nd}$ $\varepsilon_c$ | $1^{st}$ $\sigma$/GPa | $1^{st}$ $\varepsilon_c$ | $2^{nd}$ $\sigma$/GPa | $2^{nd}$ $\varepsilon_c$ |
| 8K | 26.2 | 18.4% | 22.5 | 32.2% | 20.2 | 17.4% | 24.6 | 32.2% |
| 50K | 26.3 | 19.2% | 22.8 | 33.1% | 19.5 | 18.7% | 24.3 | 33.1% |
| 100K | 25.2 | 18.2% | 23.1 | 33.3% | 18.0 | 17.3% | 22.7 | 31.4% |
| 200K | 22.8 | 16.7% | 21.0 | 31.0% | 15.4 | 16.5% | 21.7 | 30.1% |
| 300K | 18.7 | 14.5% | 20.2 | 30.1% | 13.9 | 14.9% | 22.1 | 30.5% |
| 500K | 16.1 | 12.8% | 17.7 | 28.3% | 12.1 | 13.0% | 14.1 | 24.8% |

| | P5C2 model | | | | P5C3 model | | | |
|---|---|---|---|---|---|---|---|---|
| | $1^{st}$ $\sigma$/GPa | $1^{st}$ $\varepsilon_c$ | $2^{nd}$ $\sigma$/GPa | $2^{nd}$ $\varepsilon_c$ | $1^{st}$ $\sigma$/GPa | $1^{st}$ $\varepsilon_c$ | $2^{nd}$ $\sigma$/GPa | $2^{nd}$ $\varepsilon_c$ |
| 8K | 22.4 | 19.2% | 23.3 | 37.9% | 16.3 | 19.1% | 24.9 | 37.6% |
| 50K | 21.7 | 17.8% | 23.7 | 37.7% | 15.8 | 18.0% | 25.2 | 37.4% |
| 100K | 21.3 | 17.5% | 23.7 | 37.8% | 15.5 | 17.4% | 24.6 | 36.8% |
| 200K | 20.1 | 16.4% | 21.4 | 35.9% | 14.4 | 16.0% | 23.9 | 36.4% |
| 300K | 18.9 | 15.1% | 21.7 | 36.2% | 13.4 | 14.6% | 19.2 | 32.4% |
| 500K | 15.0 | 11.6% | 17.9 | 33.1% | 11.5 | 12.0% | 19.6 | 32.9% |

| | P6C2 model | | | | P6C3 model | | | |
|---|---|---|---|---|---|---|---|---|
| | $1^{st}$ $\sigma$/GPa | $1^{st}$ $\varepsilon_c$ | $2^{nd}$ $\sigma$/GPa | $2^{nd}$ $\varepsilon_c$ | $1^{st}$ $\sigma$/GPa | $1^{st}$ $\varepsilon_c$ | $2^{nd}$ $\sigma$/GPa | $2^{nd}$ $\varepsilon_c$ |
| 8K | 22.2 | 18.7% | 23.5 | 43.3% | 16.1 | 18.5% | 23.3 | 39.6% |
| 50K | 21.8 | 18.4% | 23.5 | 43.5% | 15.8 | 18.0% | 23.8 | 40.8% |
| 100K | 21.2 | 17.6% | 17.4 | 37.8% | 15.0 | 16.8% | 23.9 | 41.3% |
| 200K | 20.0 | 16.2% | 21.7 | 41.3% | 14.4 | 15.9% | 20.5 | 38.8% |
| 300K | 17.6 | 13.9% | 21.6 | 41.8% | 13.1 | 14.2% | 19.5 | 37.6% |
| 500K | 16.1 | 12.6% | 17.9 | 38.0% | 10.6 | 11.1% | 16.0 | 33.0% |